\documentclass{aa}
\usepackage{times}
\usepackage{graphics}
\usepackage{epsf}
\begin{document}
\topmargin0.0cm
\thesaurus{06(03.20.1; 08.03.4; 08.09.2: IRC\,+10\,420; 08.13.2; 08.19.3; 08.23.2)}
\title{
The rapidly evolving hypergiant IRC\,+10\,420:
High-resolution bispectrum speckle-interferometry 
and dust-shell modelling
}
\author{
T.\ Bl\"ocker\inst{1}\and
Y.\ Balega\inst{2}\and
K.-H.\ Hofmann\inst{1}\and
J.\  Lichtenth\"aler\inst{1}\and
R. Osterbart\inst{1}\and
G. Weigelt\inst{1}
}
\institute{
Max--Planck--Institut f\"ur Radioastronomie, Auf dem H\"ugel 69,
D--53121 Bonn, Germany
\and
Special Astrophysical Observatory, Nizhnij Arkhyz, Zelenchuk region,
Karachai--Cherkesia, 35147, Russia
}
\offprints{\hbox{T.\,Bl\"ocker\,(bloecker@mpifr-bonn.mpg.de)}}
%
\date{Received date /  accepted date}
\titlerunning{The hypergiant \object{IRC\,+10\,420}:
Bispectrum speckle-interferometry and dust-shell modelling}
\authorrunning{T.\ Bl\"ocker et al.}
\maketitle
\begin{abstract}
The hypergiant \object{IRC\,+10\,420} is a unique object for the
study of stellar evolution since it is the only object that is
believed to be witnessed in its rapid transition from the  red supergiant
stage to the Wolf-Rayet phase. Its effective temperature has increased by
1000-2000\,K within only  20\,yr.
We present the first  speckle  observations
of \object{IRC\,+10\,420} with 73\,mas resolution.
A diffraction-limited  2.11\,$\mu$m image was reconstructed
from 6\,m telescope speckle data using the
bispectrum speckle-interferometry 
method. The visibility function shows
that the dust shell contributes $\sim 40\%$ to the total flux and
the unresolved central object  $\sim 60\%$.

Radiative transfer calculations have been performed
to model both the spectral energy distribution and visibility function.
The grain sizes, $a$, were found to be in accordance with
a standard distribution function,
$n(a)$\,$\sim$\,$a^{-3.5}$, with $a$ ranging between
$a_{\rm min}$\,=\,0.005\,$\mu$m and $a_{\rm max}$\,=\,0.45\,$\mu$m.
The observed dust shell properties cannot be fitted by
single-shell models but seem to require multiple components.
At a certain distance we considered an enhancement over the assumed  $1/r{^x}$
density distribution.
The best model for {\it both} SED {\it and} visibility was found  
for a dust shell with a dust temperature of 1000\,K at its inner radius
of $69\,R_{\ast}$. At a distance of
$308\,R_{\ast}$ the density was enhanced by a factor of 40 and and its
density exponent was changed from $x=2$ to $x=1.7$.
The  shell's intensity distribution was found to be ring-like.
The ring diameter
is equal to the inner diameter of the hot shell ($\sim 69$\,mas).
The diameter of the central star is $\sim 1$\,mas.
The assumption of a hotter inner shell of 1200\,K gives fits of almost
comparable quality but decreases the spatial
extension of both shells' inner boundaries
by $\sim 30$\% 
(with $x=1.5$ in the outer shell).
The two-component model  can be interpreted in terms of a termination of an
enhanced mass-loss phase roughly 60 to 90 yr (for $d=5$\,kpc) ago.
The bolometric flux, $F_{\rm bol}$, is
$8.17 \cdot 10^{-10}$\,Wm$^{-2}$ corresponding to a central-star 
luminosity of $L/L_{\odot} = 25\,462 \cdot (d/{\rm kpc})^{2}$.
%
\keywords{
Techniques: image processing ---
Circumstellar matter ---
Stars: individual: IRC\,+10\,420 ---
Stars: mass--loss ---
Stars: supergiants ---
Stars: Wolf-Rayet 
}
\end{abstract}
\section{Introduction}
The
star \object{IRC\,+10\,420}
(= \object{V\,1302~Aql} = \object{IRAS\,19244+1115}) is an
outstanding object for the study of stellar evolution.
Its spectral type changed from
F8\,I$_{\rm a}^{+}$ in 1973 (Humphreys et al.\ 1973)
to mid-A today (Oudmaijer et al. \cite{OudGroeMatBloSah96}, Klochkova et al.\
\cite{KloChePan97}) corresponding to an effective temperature increase
of 1000-2000\,K within only 20\,yr.
It is one of the
brightest IRAS objects
and one of the warmest stellar OH maser sources known
(Giguere et al.\ \cite{GigWooWeb76}, see also
Mutel et al.\  \cite{MutEtal79},
Diamond et al.\ \cite{DiaEtal83},
Bowers \cite{Bow84},
Nedoluha \& Bowers \cite{NedBow92}).
Ammonia emission has been reported by McLaren \& Betz (1980) and
Menten \& Alcolea (1995).
Large mass-loss rates, typically of the
order of several $10^{-4}$\,M$_{\odot}$/yr
(Knapp \& Morris \cite{KnaMor85},  Oudmaijer et al. \cite{OudGroeMatBloSah96})
were determined by CO\,observations.
Two evolutionary scenarios have been suggested for
\object{IRC\,+10\,420}: It
is either a post-AGB
(AGB: Asymptotic Giant Branch)
star evolving through the proto-planetary
nebula stage (e.g.\ Fix \& Cobb \cite{FixCob87}, Hrivnak et al.\
\cite{HriKwoVol89}, Bowers \& Knapp \cite{BowKnap89}),
or it is a massive hypergiant evolving from the RSG
(Red Supergiant Branch)
branch towards the Wolf-Rayet phase
(e.g.\ Mutel et al.\ \cite{MutEtal79},
Nedoluha \& Bowers \cite{NedBow92},
Jones et al.\ \cite{JonHumGehEtal93},
Oudmaijer et al.\ \cite{OudGroeMatBloSah96},
Kloch\-kova et al.\ \cite{KloChePan97}).
However, due to its distance ($d$\,=\,3-5\,kpc), 
large wind velocity (40\,km/s)
and photometric history, \object{IRC\,+10\,420} is most likely a
luminous massive star (see Jones et al.\ \cite{JonHumGehEtal93} and
Oudmaijer et al. \cite{OudGroeMatBloSah96}), therefore being the only
massive object observed until now in its transition to the
Wolf-Rayet phase. The structure of the circumstellar environment of
\object{IRC\,+10\,420} appears to be very complex  (Humphreys et al.\ 1997),
and scenarios proposed
to explain the observed spectral features of \object{IRC\,+10\,420}
include a rotating equatorial disk (Jones et al.\ 1993), 
bipolar outflows (Oudmaijer et al.\ 1994), and
the infall of circumstellar material 
(Oudmaijer 1998). 

Previous infrared speckle and coronogra\-phic
observations were reported by
Dyck et al.\ (\cite{DyckEtal84}),
Ridg\-way et al.\ (\cite{RidgEtal86}),
Cobb \& Fix (\cite{CobFix87}),
Christou et al.\ (\cite{ChrEtal90})
and Kastner \& Weintraub (\cite{KastWein95}).
In this paper we present diffraction-limited  73\,mas
bispectrum speckle-interferometry  observations
of the dust shell of \object{IRC\,+10\,420} as well as
radiative transfer calculations to model its spectral energy distribution
and visibility.
\section{Observations and data reduction}
The \object{IRC\,+10\,420} speckle interferograms
were obtained with the Russian
6\,m telescope at the Special Astrophysical Observatory
on June 13 and 14, 1998.
The speckle data were recorded
with our NICMOS-3 speckle camera
(HgCdTe array, 256$^2$ pixels,
frame rate 2 frames/s)
through an interference filter with a
centre wavelength of 2.11\,$\mu$m and a bandwidth of 0.19\,$\mu$m.
Speckle interferograms of the unresolved star HIP 95447 were taken for the
compensation of the 
speckle interferometry transfer function. The observational
parameters were as follows:
exposure time/frame 50~ms; number of frames 8400
(5200 of \object{IRC\,+10\,420} and 3200 of \object{HIP 95447});
2.11\,$\mu$m seeing (FWHM) $\sim$1\farcs0;
field of view 7\farcs8$\times$7\farcs8; pixel size 30.5\,mas.
A diffraction-limited image of  \object{IRC\,+10\,420}
with 73\,mas resolution was reconstructed from
the speckle interferograms using the bispectrum speckle-interferometry method
(Weigelt \cite{Wei77}, Lohmann et al.\ \cite{LohWeiWir83},
Hofmann \& Weigelt \cite{HofWei86}).
The bispectrum of each frame consisted of $\sim$37 million elements.
The modulus of the object Fourier transform (visibility) was
determined  with the speckle interferometry method (Labeyrie \cite{Lab70}).

It is noteworthy that 2.11\,$\mu$m filters also serve to
image hydrogen emission as, for instance, H$_{2}$ (2.125\,$\mu$m)
or Br\,$\gamma$ (2.166\,$\mu$m) emission.
Accordingly, it is possible that one might look at hydrogen emission
rather than at
the dust emission of a circumstellar shell.
The low-resolution spectrum of \object{IRC\,+10\,420} published in the
atlas of Hanson et al.\ (\cite{HanEtal96}) shows a  Br\,$\gamma$ line in
emission as the most prominent feature for the wavelength range considered
here. Oudmaijer et al.\ ({\cite{OudEtal94}) carried out
high-resolution infrared spectroscopy and found an equivalent width of
1.2\,\AA\  for the  Br\,$\gamma$ emission line. This is only 0.06\% of the
bandwidth of our interference filter and consequently negligible.

Figure~\ref{Fvisi}
shows the reconstructed 2.11\,$\mu$m visibility function of
\object{IRC\,+10\,420}. There is only marginal evidence
for an elliptical visibility shape
(position angle of the long axis $\sim 130\degr \pm 20\degr$, 
axis ratio $\sim 1.0$ to 1.1).
The visibility 0.6 at frequencies $>4$\,cycles/arcsec
shows that the stellar contribution to the total flux is $\sim$\,60\%
and the dust shell contribution is $\sim$\,40\%.
In order to compare our results
with speckle observations of other groups we determined 
the Gau\ss\ fit FWHM diameter of the dust shell to be
$d_{\rm FWHM} = (219 \pm 30)$\,mas.
By comparison, 
Christou et al.\ (1990) found for 3.8\,m telescope K-band data
a dust-shell flux contribution of $\sim$50\% and $d_{\rm FWHM} = 216$\,mas.
However, as will be shown later, 
a ring-like intensity distribution appears
to be much better suited than the assumption of a Gaussian distribution
whose corresponding FWHM diameter fit may give misleading sizes
(see Sect.~\ref{SSSint}).
Fig.~\ref{Fradial} displays the azimuthally averaged diffraction-limited
images of \object{IRC\,+10\,420} and the unresolved star HIP 95447.
\begin{figure}
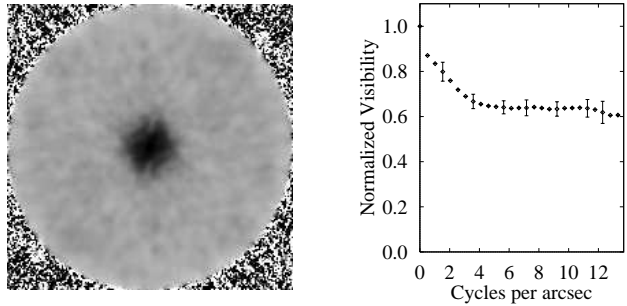

\epsfxsize=3.8cm
\parbox{4.0cm}{
\epsfbox{h1438_lt.f1}
}
\epsfxsize=4.7cm
\parbox{4.8cm}{
\epsfbox{h1438_rt.f1}
}
\caption{{\bf Left:} Two-dimensional 2.11\,$\mu$m visibility function of
\object{IRC\,+10\,420} shown up to the diffraction limit (see right panel).
The dark central structure shows that the central object is
surrounded by a dust shell.
{\bf Right:}
Azimuthally averaged  2.11\,$\mu$m visibility of \object{IRC\,+10\,420} 
with error bars for selected frequencies.
This visibility function consists of a constant plateau (visibility $\sim 0.6$)
caused by the unresolved central object  
and a triangle-shaped low-frequency
function caused by the faint extended nebula.
}
\label{Fvisi} 
\end{figure}
\begin{figure}
\epsfxsize=5.5cm
\parbox{5.5cm}{
\epsfbox{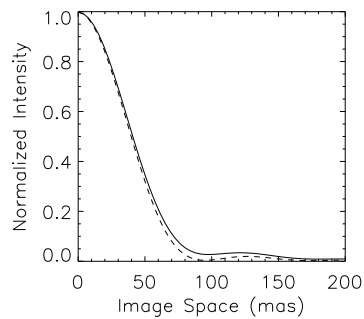}
}
\parbox{3.2cm}{
\caption{
Azimuthally averaged radial plots of the reconstructed diffraction-limited
$2.11 \mu$m-images of
\object{IRC\,+10\,420} (solid line) and \object{HIP\,95447} (dashed line).
} \label{Fradial}
}
\end{figure}
%
\section{Dust shell models}
\subsection{Spectral energy distribution}
The spectral energy distrubion (SED)
of \object{IRC\,+10\,420} with
9.7 and 18\,$\mu$m silicate emission features
is shown in Fig.~\ref{Fsedtdust}.
It corresponds to the '1992' data set used by Oudmaijer et al.\
(\cite{OudGroeMatBloSah96})
and combines VRI (October 1991),
near-infrared (March and April 1992) and Kuiper Airborne Observatory
photometry (June 1991) of Jones et al.\ (\cite{JonHumGehEtal93}) with
the IRAS measurements and 1.3\,mm data from
Walmsley et al.\ (\cite{WalEtal91}).
Additionally, we included the data of Craine et al.\
(\cite{CraEtal76}) for $\lambda < 0.55\,\mu$m.
In contrast to the near-infrared, the optical magnitudes have remained
constant during the
last twenty years within a tolerance of $\approx 0 \fm 1$.
%
%

\object{IRC\,+10\,420} is highly reddened due to an extinction of
$A^{\rm total}_{\rm V} \approx 7^{\rm m}$
by the interstellar medium and the circumstellar shell.
From polarization studies
Craine et al.\ (\cite{CraEtal76}) estimated  an interstellar extinction of
$A_{\rm V} \approx 6^{\rm m}$. Jones et al.\ (\cite{JonHumGehEtal93}) derived
from their polarization data
$A_{\rm V} \approx 6^{\rm m}$ to $7^{\rm m}$.
Based on the strength of the diffuse interstellar bands
Oudmaijer (\cite{Oud98}) inferred $E(B-V)=1 \fm 4  \pm  0 \fm 5$ for the
interstellar contribution compared to a total of  $E(B-V)=2\fm 4$.
We  will use an interstellar $A_{\rm V}$ of
$5^{\rm m}$ as in Oudmaijer et al.\ (\cite{OudGroeMatBloSah96}).
This interstellar reddening was taken into account by 
adopting the method of Savage \& Mathis (\cite{SavMat79}) with
$A_{\rm V} = 3.1 E(B-V)$.
\subsection{The radiative transfer code}
In order to model both the observed SED and $2.11\,\mu$m visibility, 
we performed radiative transfer calculations for dust shells
assuming spherical symmetry.
We used the code DUSTY developed by  Ivezi\'c et al.\ 
(\cite{IveNenEli97}), 
which solves the spherical radiative transfer problem utilizing the
self-similarity and scaling behaviour of IR emission from radiatively
heated dust (Ivezi\'c \&  Elitzur \cite{IveEli97}).
To solve the radiative transfer problem including absorption, emission and
scattering several properties of the central source and its surrounding
envelope are required, viz. (i) the spectral shape of the central source's
radiation; (ii) the dust properties,
i.e. the envelope's chemical composition and grain size distribution
as well as the dust temperature at the inner boundary; (iii) the relative
thickness of the envelope, i.e. the ratio of outer to inner shell radius,
and the  density distribution; and (iv) the total optical depth at a
given reference wavelength.
The code has been expanded for the calculation of synthetic visibilities
as described by Gauger et al.\ (\cite{GauEtal99}).

\subsection{Single-shell models}
%
We calculated various models 
considering the following parameters within the radiative transfer
calculations: SED and visibility were modelled for
$T_{\rm eff}=7000$ to 9000\,K,
black bodies and Kurucz (1992) model atmospheres
   as central sources of radiation,
different silicates (Draine \& Lee \cite{DraLee84},
   Ossenkopf et al.\ \cite{OssEtal92}, David \& Pegourie \cite{DavPeg95}), 
single-sized grains with $a=0.01$ to 0.6\,$\mu$m and 
   grain size distributions according to Mathis et al.\
  (\cite{MRN77}, hereafter MRN),
   i.e. $n(a) \sim a^{-3.5}$,
 with  0.005\,$\mu {\rm m} \leq a  \leq (0.20$ to  $0.60)$\,$\mu$m.   
We used a   $1/r^{2}$
density distribution and a shell thickness
$Y_{\rm out} = r_{\rm out}/r_{1}$ of
$10^{3}$ to $10^{5}$
with $r_{\rm out}$ and $r_{1}$ being the outer and inner radius
of the shell, respectively. Then, the
remaining fit parameters are the  dust temperature, $T_{1}$, which
determines the radius of the shell's inner boundary, $r_{1}$,
and the optical depth, $\tau$, at a given reference wavelength,
$\lambda_{\rm ref}$. We refer to 
$\lambda_{\rm ref} =  0.55\,\mu$m. Models were calculated for
dust temperatures between 400 and 1000\,K and
optical depths between 1 and 12. Significantly larger
values for $\tau$ lead to silicate features in absorption.

Fig.~\ref{Fsedtdust} shows the SED calculated for $T_{\rm eff}$=$7000$\,K,
$Y_{\rm out}$=$10^{3}$, Draine \& Lee (1984) silicates,
MRN grain size distribution ($a_{\rm max}=0.2\,\mu$m) and
different dust temperatures. It
illustrates that the long-wavelength range is sufficiently
well fitted for cool dust with  $T_{1}=400$\,K,
optical wavelengths and silicate features require
$\tau \sim 5$. 
The inner radius of the dust shell is at 
$r_{1} = 447\,R_{\ast}$ ($R_{\ast}$: stellar radius), the 
equilibrium temperature at the  outer boundary amounts to 
$T_{\rm out}= 22$\,K.
However, the fit fails in the near-infrared
%
underestimating the flux between 2 and 5\,$\mu$m. Instead this part of the
SED seems to require much hotter dust of $T_{1} \ga 800$\,K
($r_{1} \la 145\,R_{\ast}$, $T_{\rm out}=32$\,K).
This confirms the findings of Oudmaijer et al.\ (\cite{OudGroeMatBloSah96})
who conducted radiative transfer calculations in the small particle limit,
where scattering is negligible. They introduced a cool (400\,K) and a hot
(1000\,K) shell to achieve  an overall fit. 
\begin{figure}
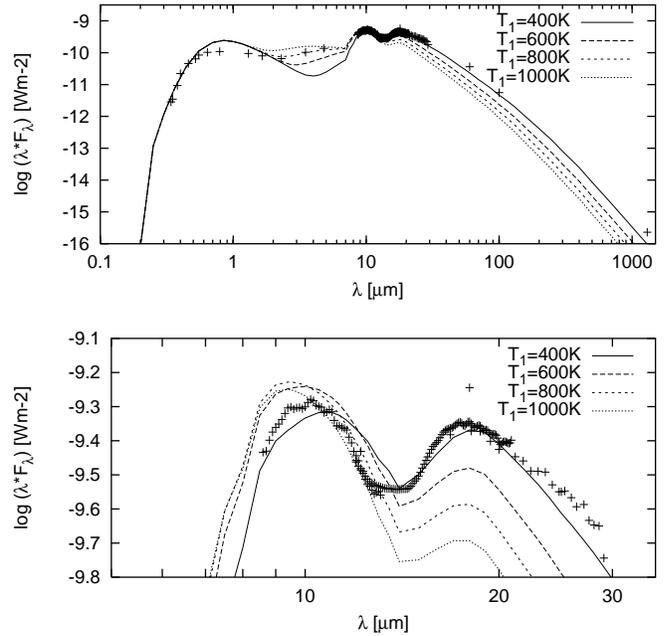

\centering
\epsfxsize=8.8cm
\mbox{\epsffile{h1438_tp.f3}}
\epsfxsize=8.8cm
\mbox{\epsffile{h1438_bt.f3}}
\protect{\vspace*{-0.2cm}}
\caption[sedtdust]
{Model SED for $T_{\rm eff}=7000$\,K, $\tau_{0.55\mu{\rm m}}=5$ and 
different dust temperatures $T_{1}$. 
The lower panel shows the  silicate features.
The calculations are based on a black body, Draine \& Lee (1984) silicates,
and an MRN grain size distribution with $a_{\rm max}=0.2\,\mu$m.
The symbols (+)
refer to the observations (see text) corrected for interstellar extinction of
$A_{\rm v}=5^{\rm m}$.
}                                      \label{Fsedtdust}
\end{figure}
\begin{figure}
\centering
\epsfxsize=8.8cm
\mbox{\epsffile{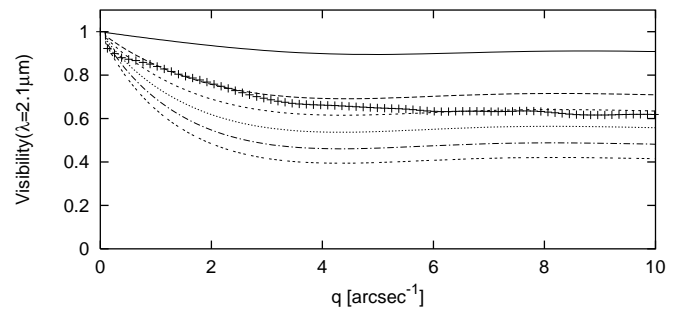}}
\caption[visiamax]
{Model visibility function for  $T_{\rm eff}=7000$\,K,
$\tau_{0.55\mu{\rm m}}=5$, $T_{1}=600$\,K and 
different maximum grain sizes in the MRN grain size distribution
($a_{\rm max}=0.2$, 0.4, 0.45, 0.5, 0.55 and $0.6\,\mu$m from top to bottom).
The calculations are based on a black body and Draine \& Lee (1984) silicates.
}                                      \label{Fvisiamax}
\protect{\vspace*{-0.4cm}}
\end{figure}
%
We found this behaviour of single-shell SEDs to be almost independent
of various input parameters.
Increasing $Y_{\rm out}$ from $10^{3}$ to $10^{5}$
leads to somewhat higher fluxes, but only
for $\lambda > 100\mu$m. 
The equilibrium temperature at the outer boundary decreases by a factor of two
if  the shell's thickness is increased  by one order of magnitude.
Larger $T_{\rm eff}$ gives slightly less flux
in the near-infrared, larger wavelengths ($\lambda > 10\mu$m) are almost
unaffected. The Draine \& Lee (1984) and David \& Pegourie (1994) silicates
give almost identical results, the optical constants of Ossenkopf et al.
(1992) lead to a larger  9.7\,$\mu$m/18\,$\mu$m flux ratio for the silicate
features, to somewhat higher fluxes between 2 and 10\,$\mu$m and to a somewhat
flatter slope of the SED at short wavelengths.
However, the need for two dust components still exists.
Calculations with different grain sizes show
that single-sized grains larger than 0.2\,$\mu$m are not suitable for
\object{IRC\,+10\,420}.
The silicate features are worse fitted and, in particular, a significant
flux deficit appears in the optical and near-infrared.
The variation of the
maximum grain size in the MRN distribution leads to much
smaller differences due to the steep decrease of the grain number density
with grain size.

The 2.11\,$\mu$m visibility 
is very sensitive against scattering, thus depending
strongly on the assumed grain sizes (see Groenewegen \cite{Groe97})
as demonstrated in Fig.~\ref{Fvisiamax}.
For a given set of parameters both {\it inclination} and {\it curvature}
of the visibility are mainly given by the optical depth, $\tau$,
and the grain size, $a$. Since $\tau$ is fixed to small values due
to the emission profiles, $a$ can be
determined.
The dust temperature must be varied simultaneously since an increase of
$T_{1}$ leads to a steeper declining visibility. 
Our calculations show that the {\em visibility}
is best fitted for an intermediate
$T_{1}=600\,K$ in contrast to the SED.
Either single-sized grains with $a \sim 0.4\,\mu$m
(which, however, are ruled out by the SED)
or MRN grain size distributions
with $a_{\rm max} \sim 0.45$ to 0.5\,$\mu$m are appropriate.
This  result still depends on the kind of silicates considered, i.e.
on the optical constants. For instance, if we take the 'warm silicates'
of Ossenkopf et al. (1992), we get somewhat smaller particles 
(by $\sim 0.1$\,$\mu$m, i.e.\ 
$a \sim 0.3$\,$\mu$m for single-sized grains 
and $a_{\rm max} \sim 0.35$\,$\mu$m for a grain distribution, resp.). 
The differences to the corresponding 'cold silicates' or to
the data from David \& Pegourie (1994) are found to be smaller.
The fits to the SED are of comparable quality.
We chose Draine \& Lee (\cite{DraLee84})  silicates with
$a_{\rm min} = 0.005\,\mu$m and $a_{\rm max}=0.45\,\mu$m.

\subsection{Multiple dust-shell compenents}
\subsubsection{Two component shells} \label{SSmultdust}
Since we failed to model the SED with the assumptions made so far,
we introduced a two-component shell as Oudmaijer et al.\ (1996).
For that purpose, we assume
that \object{IRC\,+10\,420}
had passed through a superwind phase in its history as
can be expected from its evolutionary status 
(see Schaller et al.\ 1992, Garc\'{i}a-Segura et al.\ 1996).
This is in line with
the conclusions drawn from  the  Oudmaijer et al.\ (1996) model 
and recent interpretations of HST data (Humphreys et al.\ 1997).
A previous superwind phase leads to changes in the density distribution, i.e.\
there is a region in the dust shell
which shows a density enhancement over the normal
$1/r^{2}$ distribution. The radial density distribution may also change
within this superwind shell. For more details, see Suh \& Jones
(\cite{SuhJon97}). Since dust formation operates on very short timescales
in OH/IR stars, we assume a constant outflow velocity for most of the
superwind phase and thus a $1/r^{2}$ density distribution. 
For simplicity, we consider only single jumps with enhancement factors, or
amplitudes, $A$
at radii $Y=r/r_{1}$ in the relative density distribution as demonstrated
in Fig.~\ref{Fswmodel}.

Concerning the grains we stay with Draine \& Lee (\cite{DraLee84}) silicates
and an MRN grain size distribution with $a_{\rm min} = 0.005\,\mu$m and
$a_{\rm max}=0.45\,\mu$m as in the case of the single shell models. The
influence of different grain-size distributions will be discussed later.
%
\begin{figure}
\centering
\epsfxsize=4.4cm
\parbox{4.6cm}{
\mbox{\epsffile{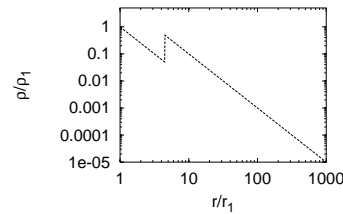}}
}
\parbox{4.0cm}{
\caption[swmodel]{
Relative density distribution for a superwind at $Y=r/r_{1}=4.5$ with
an amplitude of $A=10$.} \label{Fswmodel}
}
\end{figure}

We calculated a grid of models for $T_{1}=1000$\,K
with superwinds at $Y = 2.5$ to 8.5
with amplitudes $A$ ranging from 10 to 80. Due to the introduced
density discontinuity the flux conservation has
to be controlled carefully, 
in particular at larger optical depths and amplitudes.
SED and visibility behave contrarily concerning the adjustment of the
superwind: The SED requires sufficiently large distances,
$Y \ga 4.5$, and moderate amplitudes, $A \la 20$ to 40, in particular for 
the flux between 2 and $10\,\mu$m and  for $\lambda > 20\mu$m.
\begin{figure}
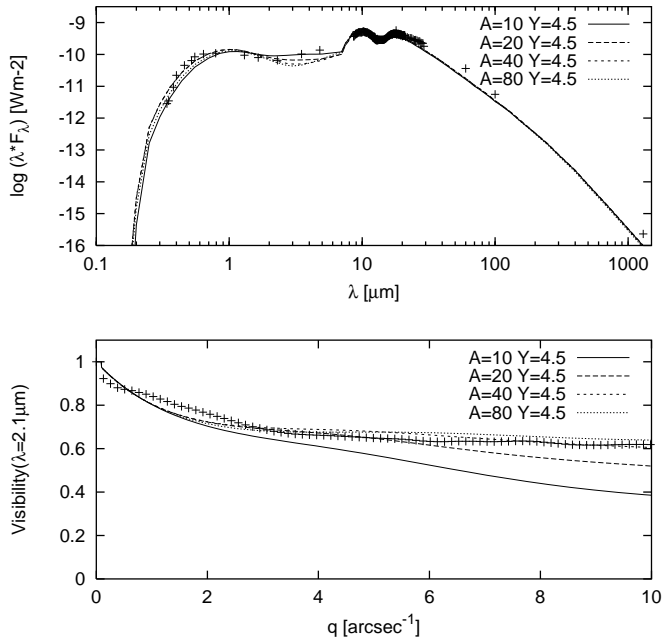

\centering
\epsfxsize=8.8cm
\centering
\mbox{\epsffile{h1438_tp.f6}}
\centering
\epsfxsize=8.8cm
\mbox{\epsffile{h1438_bt.f6}}
\caption[Fsedvisisw]
{
SED (top) and visibility (bottom)
for a superwind model with  $Y=r/r_{1}=4.5$ and different amplitudes.
Model parameters are:
black body, $T_{\rm eff}=7000$\,K, $T_{1}=1000$\,K,
$\tau_{0.55\mu{\rm m}}=7.0$, Draine \& Lee (1984) silicates,
Mathis et al.\ (1977) grain size distribution with $a_{\rm max}=0.45\,\mu$m,
and $Y_{\rm out}=10^{4}$.
The symbols
refer to the observations (see text) corrected for interstellar extinction of
$A_{\rm v}=5^{\rm m}$.
}                                      \label{Fsedvisisw}
\end{figure}
A good fit was found for $Y=6.5$ and $A=20$ corresponding to
$r_{1} = 81\,R_{\ast}$. 
Note that the bolometric flux at the inner dust-shell radius
(and therefore $r_{1}/r_{\ast}$) is
fully determined by the solution of the radiative transfer problem
even though the overall luminosity is not (Ivezi\'c \& Elitzur 1997).
The dust temperature
at the density enhancement ($r_{2}= 527\,R_{\ast}$) has dropped to 322\,K. 
This agrees well with the model of
Oudmaijer et al.\ (\cite{OudGroeMatBloSah96}).
The visibility, however, behaves differently. In order to reproduce the
unresolved component (the plateau) large amplitudes, $A \ga 40$ to 80,
are required.
On the other hand, the slope at low spatial frequencies is best reproduced
for a close superwind shell, $Y < 4.5$ (at this distance independent on $A$).
The best model found for {\it both} SED {\it and} visibility 
is that with $Y=4.5$ and $A=40$ as shown in Fig.~\ref{Fsedvisisw}.
It corresponds to $r_{1}= 71\,R_{\ast}$ and
$r_{2}= 320\,R_{\ast}$ (with $T_{2} \sim 475$\,K), i.e.\ to
angular diameters of $\Theta_{1} = 71$\,mas  
and $\Theta_{2} = 321$\,mas. The angular diameters depend on the
model's bolometric flux, $F_{\rm bol}$, which is
$8.17 \cdot 10^{-10}$\,Wm$^{-2}$. Accordingly, the central star
has a luminosity of $L/L_{\odot} = 25\,462 \cdot (d/{\rm kpc})^{2}$ and
an angular diameter of
$\Theta_{\ast} = 1.74 \cdot 10^{9} \,
                 \sqrt{ F_{\rm bol}/T_{\rm eff}^{4} } \sim 1$\,mas.
Assuming a constant outflow velocity of $v = 40$\,km/s,
the expansion ages of
the two components are $t_{1}/{\rm yr} = 4.2 \cdot (d/{\rm kpc})$ and
$t_{2}/{\rm yr} = 18.9 \cdot (d/{\rm kpc})$.
With a dust-to-gas ratio of 0.005 and a
specific dust density of 3\,g\,cm$^{-3}$ the mass-loss rates
of the components are
$\dot{M}_{1}= 1.4 \cdot 10^{-5}$\,$M_{\odot}/{\rm yr}\cdot (d/{\rm kpc})$
and
$\dot{M}_{2}= 5.5 \cdot 10^{-4}$\,$M_{\odot}/{\rm yr}\cdot (d/{\rm kpc})$.

Fig.~\ref{Fffraction} shows the fractional contributions of the direct stellar
radiation, the scattered radiation and the dust emission
to the total emerging flux.
The stellar contribution has its
maximum at 2.2\,$\mu$m where it contributes 60.4\% to the total flux in
accordance with the observed visibility plateau of 0.6. 
At  this wavelength scattered radiation and dust emission amount to 25.6\%
and 14\% of the total flux, respectively. Accordingly, 64.6\% of the
2.11\,$\mu$m dust-shell emission is due to scattered stellar light and
35.3\% due to direct thermal emission from dust.
For $\lambda \la 1\,\mu$m the flux is determined by scattered radiation
whereas for  $\lambda \ga 10\,\mu$m dust emission dominates completely.
\begin{figure}
\centering
\epsfxsize=8.8cm
\centering
\mbox{\epsffile{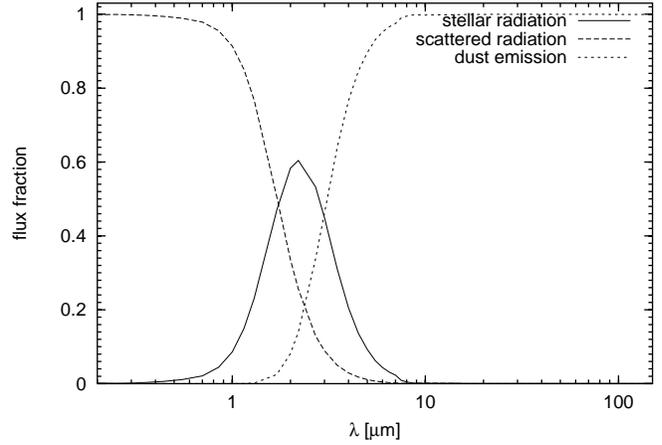}}
\caption[ffraction]
{Fractional contributions of the emerging stellar radiation 
 as well as  of the scattered radiation and of the dust emission to the total
 flux as a function of the wavelength for a superwind model with
 $Y=r/r_{1}=4.5$ and $A=40$.
 Model parameters are:
 black body, $T_{\rm eff}=7000$\,K, $T_{1}=1000$\,K,
 $\tau_{0.55\mu{\rm m}}=7.0$, Draine \& Lee (1984) silicates,
 Mathis et al.\ (1977) grain size distribution with $a_{\rm max}=0.45\,\mu$m.
}                                      \label{Fffraction}
\end{figure}
%
\subsubsection{Influence of the grain-size distribution}
As in the case of the single-shell models we also studied other
grain size distributions. The MRN distribution derived for the interstellar
medium gives a continuous decrease of the number density with increasing
grain sizes. On the other hand, the distribution of grains in dust-shells of
evolved stars rather appears to be peaked at a dominant size
(e.g.\ Kr\"uger \& Sedlmayr 1997, Winters et al.\ 1997).
It is noteworthy that even in the case of a sharply peaked size distribution
the few larger particles can contribute significantly to the absorption and
scattering coefficients (see Winters et al.\ 1997). Accordingly, the
2.11\,$\mu$m visibility reacts sensitively if some larger particles are added
whereas the SED does not, as demonstrated in the previous section.
In order to study the influence of different grain size distributions on the
two-component model we calculated grids of models with
$n(a) \sim a^{q}$ for different exponents ($q=-3.0$ to $-5.5$) and lower
and upper cut-offs ($a_{\rm min}=0.005$ to 0.05\,$\mu$m and
$a_{\rm max} = 0.1$ to 0.8\,$\mu$m). Additionally we considered single-sized
grains ($a=0.1$ to 0.8\,$\mu$m).

Concerning the visibility, a larger (smaller) negative exponent in the
distribution function can, in principle, be
compensated by increasing (decreasing) the maximum grain size. For instance,
$q=-4.0$ requires $a_{\rm max} =0.55\,\mu$m to fit the 2.11\,$\mu$m visibility.
On the other hand, if the distribution becomes too narrow, the SED cannot
be fitted any longer since the $9.7\,\mu$m silicate feature turns into
absorption. A distribution with $q=-3.8$ and
$a_{\rm max}=0.50\,\mu$m best reproduces the flux-peak ratio of the silicate
features.

\begin{figure}
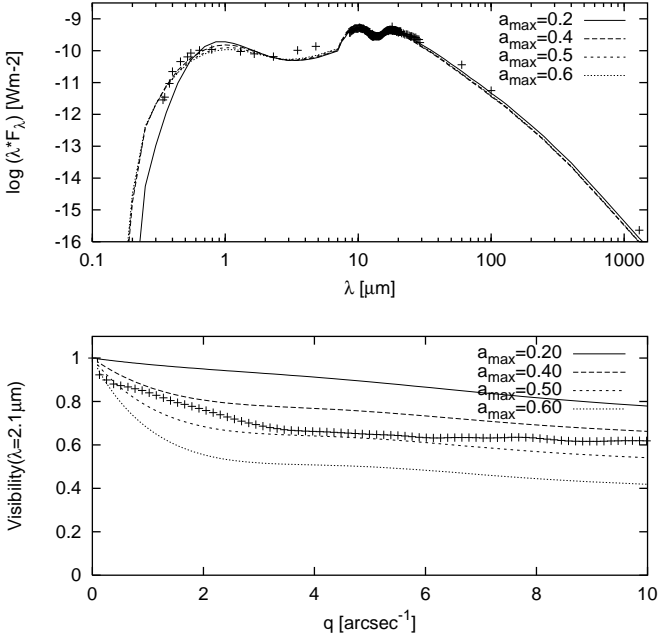

\centering
\epsfxsize=8.8cm
\centering
\mbox{\epsffile{h1438_tp.f8}}
\centering
\epsfxsize=8.8cm
\mbox{\epsffile{h1438_bt.f8}}
\caption[Fsedvisigdmax]
{
SED (top) and visibility (bottom)
for a superwind model with  $Y=r/r_{1}=4.5$ and $A=40$
calculated for
Mathis et al.\ (1977) grain size distributions with $a_{\rm max}=0.20$, 0.4,
0.5 and 0.6\,$\mu$m.
Model parameters are:
black body, $T_{\rm eff}=7000$\,K, $T_{1}=1000$\,K,
$\tau_{0.55\mu{\rm m}}=7.3$, Draine \& Lee (1984) silicates,
and $Y_{\rm out}=10^{4}$.
The symbols
refer to the observations (see text) corrected for interstellar extinction of
$A_{\rm v}=5^{\rm m}$.
}                                      \label{Fsedvisigdmax}
\end{figure}
For a given exponent in the grain-size distribution function of $q=-3.5$
we arrive at the same maximum grain
size as in the case of the one-component model, viz. 0.45\,$\mu$m, in order
to yield a fit for both the SED and the  visibility
(see Fig.~\ref{Fsedvisigdmax}). This is due to the fact
that larger particles increase the curvature of the visibility curve at
low spatial frequencies whereas the  high-frequency tail (the
plateau) is found at lower visibility values. On the other hand, the inclusion
of some larger particles does not change the shape of the SED
as discussed above.

If sufficiently small, the lower cut-off grain size can be changed
moderately (within a factor of two) without any significant change for
SED and visibility. If $a_{\rm min}$ 
exceeds, 
say, $0.05 \mu$m, the
fits of the observations  begin to become worse. For instance, the
curvature of the visibility at low spatial frequencies and the flux-peak
ratio of the silicate features are then overestimated.

Finally, we repeated the calculations under the assumption of
single-sized grains. In order to model the visibility a grain size $a$ close
to $0.3\,\mu$m is required as shown in Fig.~\ref{Fsedvisigrain}.
In contrast, the reproduction of the relative strengths of
the silicate features seems to require smaller grains,
viz. close to $0.1\,\mu$m.
Consequently, for the modelling of \object{IRC\,+10\,420}
a grain size distribution appears to be much better suited than single-sized
grains. 
\begin{figure}
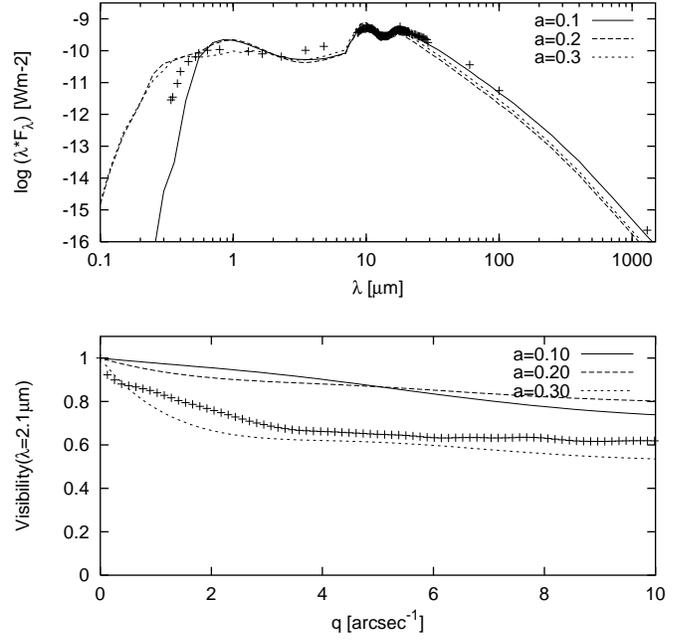

\centering
\epsfxsize=8.8cm
\centering
\mbox{\epsffile{h1438_tp.f9}}
\centering
\epsfxsize=8.8cm
\mbox{\epsffile{h1438_bt.f9}}
\caption[Fsedvisigrain]
{
SED (top) and visibility (bottom)
for a superwind model with  $Y=r/r_{1}=4.5$ and $A=40$ for single-sized grains
with  $a=0.1$, 0.2 and 0.3\,$\mu$m.
Model parameters are:
black body, $T_{\rm eff}=7000$\,K, $T_{1}=1000$\,K,
$\tau_{0.55\mu{\rm m}}=7.3$, Draine \& Lee (1984) silicates,
and $Y_{\rm out}=10^{4}$.
The symbols
refer to the observations (see text) corrected for interstellar extinction of
$A_{\rm v}=5^{\rm m}$.
}                                      \label{Fsedvisigrain}
\end{figure}
\subsubsection{Influence of the density distribution}
Inspection of the best fits derived so far reveals that there are still some
shortcomings of the models.
First, although being within the observational error bars,
the model visibilities always show a larger curvature at low spatial
frequencies. This seems to be almost independent of the chosen grain-size
distribution. Second, the flux beyond $20\,\mu$m is somewhat too low.
This may be due to our choice of a $1/r^{2}$ density distribution for both
shells. We recalculated the model grid for different $1/r^{x}$ density
distributions for both shells with x ranging between 1 and 4.
A flatter distribution  in the outer shell increases the flux in the
long-wavelength range as required but leads also to a drop of the flux in the
near-infrared. The plateau in the visibility curve remains unaffected but the
curvature at low spatial frequencies is increased.
To take advantage of the better far-infrared properties of cool shells with
flatter density distributions, but to counteract their disadvantage in the
near-infrared and at low spatial frequencies, the density distribution of the
inner shell also has to be changed. It should be somewhat steeper than the
normal $1/r^{2}$ distribution. Then the near-infrared flux is  raised and the 
visibility shows a smaller curvature in the low-frequency range. It should
be noted that the curvature is most affected for superwinds of low amplitudes. 
However, the steeper density decrease in the inner shell leads to increasingly
low visibility values in the high frequency range. Since this has to be
compensated by an increase of the superwind amplitude the advantages of the
steeper distribution are almost cancelled. 

\begin{figure}
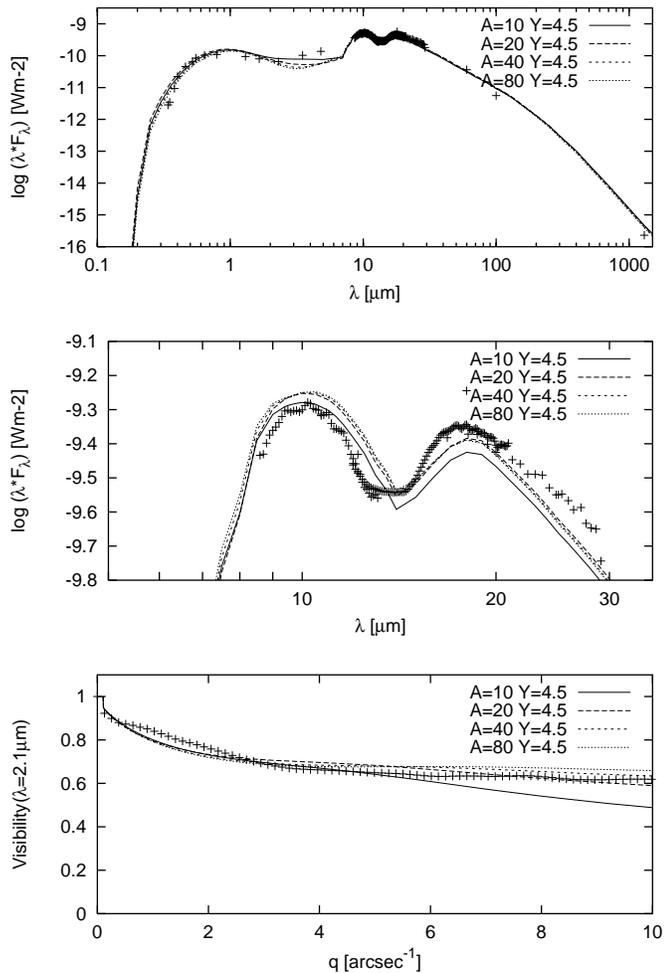

\centering
\epsfxsize=8.8cm
\centering
\mbox{\epsffile{h1438_tp.f10}}
\centering
\epsfxsize=8.8cm
\mbox{\epsffile{h1438_md.f10}}
\centering
\epsfxsize=8.8cm
\mbox{\epsffile{h1438_bt.f10}}
\caption[Fsedvisidens17]
{
SED (top),
silicate features (middle)
and visibility (bottom)
for a superwind model with  $Y=r/r_{1}=4.5$ and different amplitudes. The
inner shell obeys a $1/r^{2}$ density distribution, the outer shell a
$1/r^{1.7}$ density distribution.
Model parameters are:
black body, $T_{\rm eff}=7000$\,K, $T_{1}=1000$\,K,
$\tau_{0.55\mu{\rm m}}=7.0$, Draine \& Lee (1984) silicates,
Mathis et al.\ (1977) grain size distribution with
$a_{\rm max}=0.45\,\mu$m, and $Y_{\rm out}=10^{4}$.
The symbols
refer to the observations (see text) corrected for interstellar extinction of
$A_{\rm v}=5^{\rm m}$.
}                                      \label{Fsedvisidens17}
\end{figure}
Thus, we can stay with  a $1/r^{2}$ density distrubution in the inner
shell and moderate superwind amplitudes ($A \sim 40$). The then
best suited models we found are those with superwinds at $Y=4.5$ and 
a  $1/r^{1.7}$ distribution in the outer shell.
The corresponding SED and visibility are shown in Fig.~\ref{Fsedvisidens17} for
different superwind amplitudes. We note again that
the quality of the fits is in particular determined by the outer shell,
whereas the inner shell's exponent is less constrained.
A $1/r^{3}$ distribution in the inner shell and large superwind amplitudes
($A  \ga 80$) give similar results.

The radii of the inner and outer shell are $r_{1}= 69\,R_{\ast}$ and
$r_{2}= 308\,R_{\ast}$ (with $T_{2} \sim 483$\,K), resp., corresponding 
to angular diameters of $\Theta_{1} = 69$\,mas  
and $\Theta_{2} = 311$\,mas.
Adopting the same assumptions for outflow velocity, dust-to-gas ratio
and specific dust density as in the previous section,
the expansion ages are
$t_{1}/{\rm yr} = 4.1 \cdot (d/{\rm kpc})$ and
$t_{2}/{\rm yr} = 18.4 \cdot (d/{\rm kpc})$, for the mass-loss rate of
the inner component one gets 
$\dot{M}_{1}= 1.35 \cdot 10^{-5}$\,$M_{\odot}/{\rm yr}\cdot (d/{\rm kpc})$.
In the outer component either the outflow velocity has increased or
the mass-loss rate has decreased with time due to the more shallow density
distribution.
Provided the outflow velocity has kept constant,
the mass-loss rate at the end of the superwind phase, 92\,yr ago, was
$\dot{M}_{2} = 5.4 \cdot 10^{-4}$\,$M_{\odot}/{\rm yr}\cdot (d/{\rm kpc})$,
and, for instance, amounted to   
$\dot{M}_{2} = 8.0 \cdot 10^{-4}$\,$M_{\odot}/{\rm yr}\cdot (d/{\rm kpc})$
200\,yr ago.

Since the flatter density distribution provides a better fit for the
long-wavelength range of the SED, while the visibility is equally well fitted
compared to the standard density distribution, it is superior to the model
of 
Sect.~\ref{SSmultdust}.
Fig.~\ref{Fffraction17} gives the fractional
flux contributions (stellar, dust, scattering) for the same model as shown
in Fig.~\ref{Fffraction} but with an $1/r^{1.7}$ distribution in the outer 
shell.
The various flux contributions at 2.11\,$\mu$m are very similar to  those 
of the $1/r^{2}$ model: 62.2\% stellar light, 26.1\% scattered radiation and 
10.7\% dust emission. Thus, the total emission of the circumstellar shell is 
composed of 70.9\%  scattered stellar light and 29.1\% direct thermal 
emission from dust.
\begin{figure}
\centering
\epsfxsize=8.8cm
\centering
\mbox{\epsffile{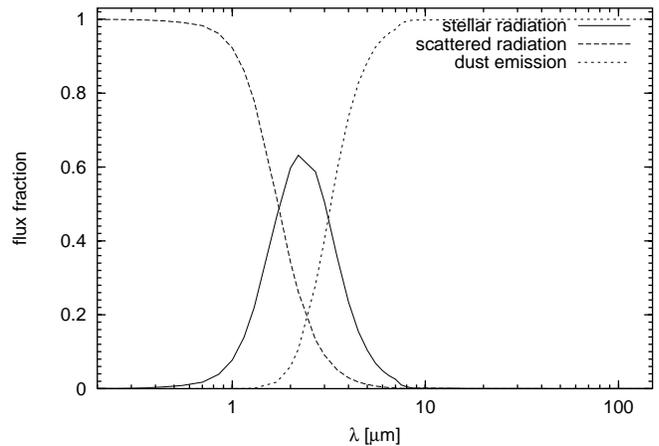}}
\caption[ffraction17]
{Fractional contributions of the emerging stellar radiation 
 as well as  of the scattered radiation and of the dust emission to the total
 flux as a function of the wavelength for a superwind model with
 $Y=r/r_{1}=4.5$, $A=40$ and a $1/r^{1.7}$ density distribution in the outer
 shell.
 Model parameters are:
 black body, $T_{\rm eff}=7000$\,K, $T_{1}=1000$\,K,
 $\tau_{0.55\mu{\rm m}}=7.0$, Draine \& Lee (1984) silicates,
 Mathis et al.\ (1977) grain size distribution with $a_{\rm max}=0.45\,\mu$m.
}                                      \label{Fffraction17}
\end{figure}
%
\subsubsection{Influence of the dust temperature}
Finally, we studied the influence of the dust temperature at the inner
boundary of the hot shell. For that purpose we recalculated the previous model
grids for dust temperatures of 800 and 1200\,K. As already shown for the 
single-shell models, an increase of the temperature at the inner boundary 
increases the flux in the near-infrared and substantially lowers the flux in
the long-wavelength range. 
On the other hand, the higher the temperature the less is the curvature of the 
visibility at low spatial frequencies, the plateau is only significantly
affected for low-amplitude superwinds. The shape of SED and 2.11\,$\mu$m
visibility for different dust temperatures at the hot shell's inner boundary
for a given superwind is demonstrated in Fig.~\ref{Fsedvisitdust}.
A temperature less than 1000\,K can be excluded in particular due to the
worse fit of the visibility for low frequencies.
Instead, the 1200\,K model gives 
a much better fit to the visibility than previous ones. 
Fig.~\ref{Fsedvisitdust} refers to an amplitude of $A=40$ in order to be 
directly comparable with the models shown before. We note that we get an
even better fit assuming $A=80$, which leaves the low-frequency-range unchanged
but improves the agreement with the measured plateau.
\begin{figure}
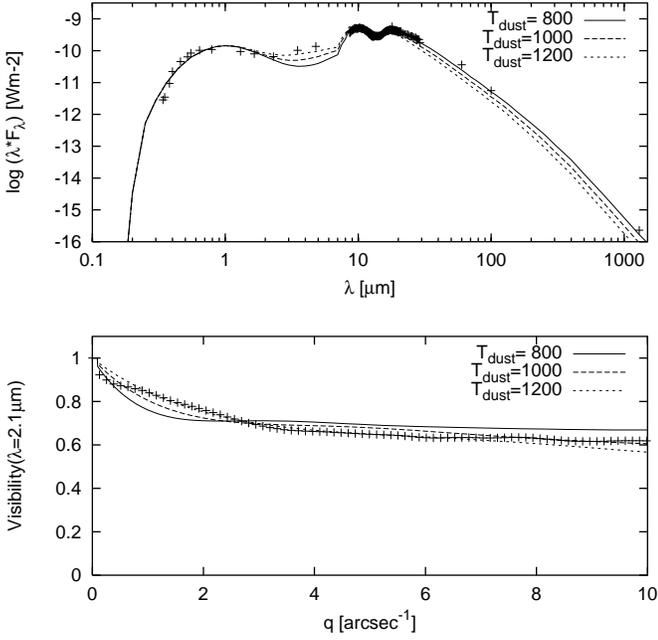

\centering
\epsfxsize=8.8cm
\centering
\mbox{\epsffile{h1438_tp.f12}}
\centering
\epsfxsize=8.8cm
\mbox{\epsffile{h1438_bt.f12}}
\caption[Fsedvisitdust]
{
SED (top) and visibility (bottom) 
for a superwind model ($Y=r/r_{1}=4.5$ and $A=40$)
with different temperatures for the inner boundary of
the hot shell. 
Model parameters are:
black body, $T_{\rm eff}=7000$\,K, 
$\tau_{0.55\mu{\rm m}}=7.0$, Draine \& Lee (1984) silicates,
Mathis et al.\ (1977) grain size distribution with
$a_{\rm max}=0.45\,\mu$m and $Y_{\rm out}=10^{4}$.
The symbols
refer to the observations (see text) corrected for interstellar extinction of
$A_{\rm v}=5^{\rm m}$.
}                                      \label{Fsedvisitdust}
\end{figure}
However, the improvement of the 2.11\,$\mu$m visibility model due 
to a hotter inner  shell with  $T_{1} = 1200$\,K
is at the expense of a considerable amplification
of the flux deficit for $\lambda \ga 20\,\mu$m  in the SED.
In order to compensate this effect we have had to assume a flatter density
profile for the outer shell than in the case of the $T_{1}=1000$\,K.,
viz.\ $\sim 1/r^{1.5}$ instead of  $\sim 1/r^{1.7}$. The corresponding
curves are shown in Fig.~\ref{Fsedvisi1200dens15}. Again, increasing the
far-infrared fluxes, as required to model the SED, leads to an increase of
the  2.11\,$\mu$m visibility's curvature at low spatial frequencies
giving somewhat worse fits for the visibility. We note that the peak-ratio
of the silicate features is better matched with a lower dust temperature of
$T_{1} = 1000$\,K. 
\begin{figure}
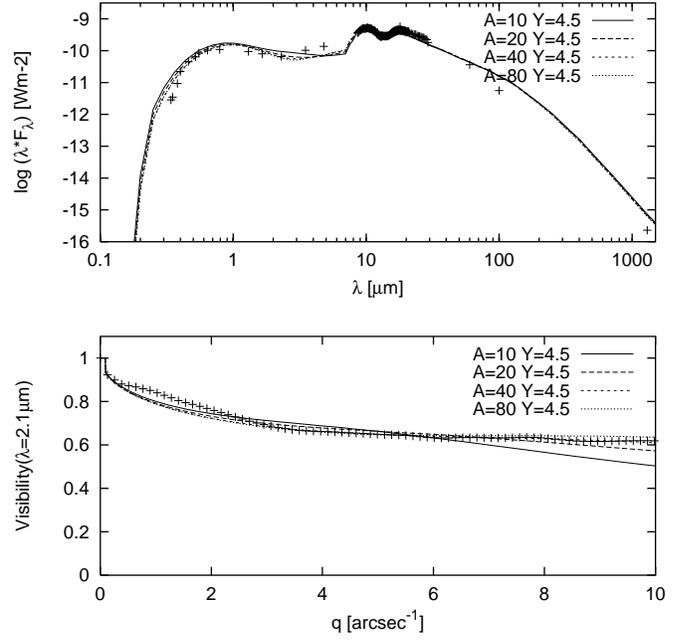

\centering
\epsfxsize=8.8cm
\centering
\mbox{\epsffile{h1438_tp.f13}}
\centering
\epsfxsize=8.8cm
\mbox{\epsffile{h1438_bt.f13}}
\caption[Fsedvisi1200dens15]
{
SED (top) and visibility (bottom)
for a superwind model with  $Y=r/r_{1}=4.5$ and different amplitudes. The
inner shell obeys a $1/r^{2}$ density distribution, the outer shell a
$1/r^{1.5}$ density distribution. The temperature at the inner boundary  of
the hot shell is 1200\,K.
Model parameters are:
black body, $T_{\rm eff}=7000$\,K,
$\tau_{0.55\mu{\rm m}}=7.0$, Draine \& Lee (1984) silicates,
Mathis et al.\ (1977) grain size distribution with
$a_{\rm max}=0.45\,\mu$m, and $Y_{\rm out} = 10^{4}$.
The symbols
refer to the observations (see text) corrected for interstellar extinction of
$A_{\rm v}=5^{\rm m}$.
}                                      \label{Fsedvisi1200dens15}
\end{figure}
 
The radii of the inner and outer shell are now considerably smaller than 
those of the previous models due to the higher temperature of the hot shell.
The radiative transfer calculations give here 
$r_{1}= 47\,R_{\ast}$ and
$r_{2}= 210\,R_{\ast}$ (with $T_{2} \sim 594$\,K), resp., resulting in 
angular diameters of $\Theta_{1} = 47$\,mas  
and $\Theta_{2} = 212$\,mas.
Accordingly,
the expansion ages are
$t_{1}/{\rm yr} = 2.8 \cdot (d/{\rm kpc})$ and
$t_{2}/{\rm yr} = 12.6 \cdot (d/{\rm kpc})$,
for the mass-loss rate of
the inner component one gets 
$\dot{M}_{1}= 9.2 \cdot 10^{-6}$\,$M_{\odot}/{\rm yr}\cdot (d/{\rm kpc})$.
Provided the outflow velocity has kept constant,
the mass-loss rate at end of the superwind phase, 63\,yr ago, was
$\dot{M}_{2} = 3.7 \cdot 10^{-4}$\,$M_{\odot}/{\rm yr}\cdot (d/{\rm kpc})$,
and, for instance, amounted to   
$\dot{M}_{2} = 5.4 \cdot 10^{-4}$\,$M_{\odot}/{\rm yr}\cdot (d/{\rm kpc})$
200\,yr ago.
%
\subsubsection{Intensity distributions} \label{SSSint}
Fig.~\ref{Fintensity17} displays the spatial distribution of the
obtained normalized
model intensity for the model shown in Fig.~\ref{Fsedvisidens17}
($T_{1} = 1000$\,K, $Y=4.5$, $A=40$, $1/r^{2}$ and $1/r^{1.7}$ density
distribution in the inner and outer shell, resp.)
The  (unresolved) central peak belongs to the central star, and the two
local intensity maxima to 
the loci of the inner rims of the two shells at 35\,mas and 157\,mas, resp.
The  $2.11\,\mu$m intensity shows a ring-like distribution with 
a steep decline with increasing distance from the inner
boundary of the circumstellar shell.
Similarly shaped intensity distributions have also been found by
Ivezi\'c \& Elitzur (1996) for optically thin shells.
\begin{figure}
\centering
\epsfxsize=8.8cm
\centering
\mbox{\epsffile{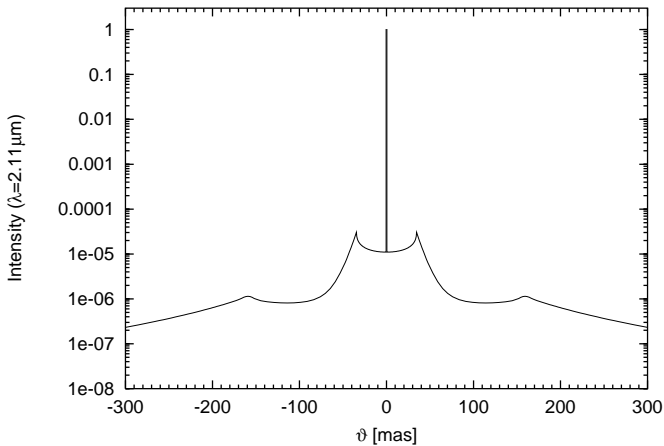}}
\caption[intensity17]
{Normalized intensity vs.\ angular displacement $\vartheta$
 for a superwind model with
 $Y=r/r_{1}=4.5$, $A=40$ and a $1/r^{1.7}$ density distribution in the outer
 shell. The (unresolved) central peak belongs to the central star.
 The inner hot rim of the circumstellar shell has a radius of 35\,mas, and
 the cool component is located at 155\,mas. Both loci correspond to local
 intensity maxima. 
 Model parameters are:
 black body, $T_{\rm eff}=7000$\,K, $T_{1}=1000$\,K,
 $\tau_{0.55\mu{\rm m}}=7.0$, Draine \& Lee (1984) silicates,
 Mathis et al.\ (1977) grain size distribution with $a_{\rm max}=0.45\,\mu$m.
}                                      \label{Fintensity17}
\end{figure}

We recall that this intensity distribution is based on radiative transfer
models taking into account both the SED and the  $2.11\,\mu$m visibility.
Figure~\ref{Fvisidens17_long} shows the model visibilities for much higher
spatial frequencies than covered by the present observations. The required
baselines would correpond to $\sim 22$ and 440\,m instead to 6\,m
(13.6\,cycles/arcsec). Since the dust-shell's diameter is $\sim 70$\,mas
a plateau is only reached for spatial frequencies larger than,
say, 15 cycles/arcsec depending on the strength of the superwind.
The central star is resolved at spatial
frequencies of $\sim 1000$ cycles/arcsec.
At frequencies $\la 2$\,cycles/arcsec the shape of the observed and
the modelled visibility function is triangle-shaped, which is a consequence
of the ring-like intensity distribution of the dust shell.
%
\begin{figure}
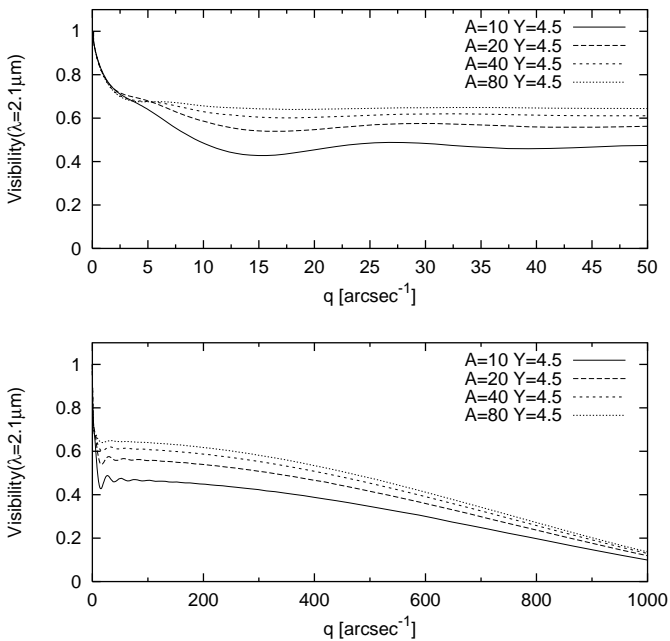

\centering
\epsfxsize=8.8cm
\mbox{\epsffile{h1438_tp.f15}}
\centering
\epsfxsize=8.8cm
\mbox{\epsffile{h1438_bt.f15}}
\caption[visidens17_long]
{Model visibility up to 50 (top) and
1000 cycles/arcsec (bottom)
for a superwind model with  $Y=r/r_{1}=4.5$ and different amplitudes. The
inner shell obeys a $1/r^{2}$ density distribution, the outer shell a
$1/r^{1.7}$ density distribution.
Model parameters are:
black body, $T_{\rm eff}=7000$\,K, $T_{1}=1000$\,K,
$\tau_{0.55\mu{\rm m}}=7.0$, Draine \& Lee (1984) silicates,
and Mathis et al.\ (1977) grain size distribution with
$a_{\rm max}=0.45\,\mu$m.
}                                      \label{Fvisidens17_long}
\end{figure}

Visibility observations are often characterized by fits with
Gaussian intensity distributions.
The resulting Gaussian FWHM diameter is then assumed to give
roughly the
typical size of the dust shell.
A Gau\ss\  fit to the observed visibility would yield a
FWHM dust-shell diameter of
(219 $\pm$ 30)\,mas  in agreement with the one given by
Christou et al.\ (1990).
However, radiative transfer models show that
a ring-like  intensity distributions appears to be more appropriate than
a Gaussian one for the dust shell of \object{IRC\,+10\,420}.
The distribution shows
a limb-brightenend dust condensation zone and a ring  diameter of 70\,mas.
\section{Summary}
Radiative transfer calculations show that the near-infrared visibility
strongly constrains
dust shell models since it is, e.g., a sensitive indicator of the grain size.
Accordingly, high-resolution interferometry results provide essential
ingredients for models of circumstellar dust-shells. 
Assuming spherical symmetry we carried out radiative transfer calculations
for the hypergiant \object{IRC\,+10\,420} to model both its SED and 
$2.11\,\mu$m visibility.
Since we failed to find good SED fits 
for single-component models,
we improved our density distribution introducing 
a second component with enhanced values
at a certain distance.
For different scaled distances $Y=r/r_{1}$ and density enhancements $A$
of this cool 
component we considered different grain-size distributions
$n(a) \sim a^{q}$,
density distributions $\rho \sim 1/r^{x}$ within the shells, and 
temperatures $T_{1}$ at the inner boundary of the hot shell.

An MRN grain size
distribution $n(a) \sim a^{-3.5}$
with $0.005\,\mu{\rm m} \leq a  \leq  0.45\mu{\rm m}$ was found to be well
suited for \object{IRC\,+10\,420}. Larger negative exponents, i.e.\ a 
narrower 
distribution, can be accounted for by increasing the maximum
grain size. For instance,  $n(a) \sim a^{-3.8}$ requires 
$a_{\rm max} \sim 0.55\,\mu$m. However, 
the range of appropriate exponents seemed to be quite small and steeper
declining distributions led to significantly worse fits. 

Assuming a $1/r^{2}$ density distribution for both shells and $T_{1}=1000\,$K
gives the best fit for $Y=4.5$ and $A=40$ (Fig.~\ref{Fsedvisisw}).
This model can be improved by 
introducing a somewhat flatter density distribution, viz.\ $\sim 1/r^{1.7}$,
for the outer shell leading to a better match with the observed SED for 
$\lambda \ga 20\,\mu$m. The quality of the visibility fit remains almost
unchanged (Fig.~\ref{Fsedvisidens17}).
Both models show a somewhat larger curvature of the visibility
at low spatial frequencies. However, the deviations are within the 
observational uncertainties.
The various flux contributions at 2.11\,$\mu$m  
are 62.2\% stellar light, 26.1\% scattered radiation and 
10.7\% dust emission.

Alternatively one may increase 
the temperature at the inner boundary of the hot shell to $T_{1}=1200\,$K
which gives somewhat better matches to the near-infrared flux and lowers the
low-frequency visibility curvature. 
To counteract the concomitant loss of flux in the far-infrared
one has to assume a  $1/r^{1.5}$ density distribution
(Fig.~\ref{Fsedvisi1200dens15}).
The fit to the silicate features is, however,
somewhat worse than in the case of the $T_{1}=1000\,$K model.

The intensity distribution was found  to be ring-like.
This appears to be typical for optically thin shells
(here $\tau_{0.55\mu{\rm m}}=7$, $\tau_{2.11\mu{\rm m}}=0.55$; see also
Ivezi\'c \& Elitzur 1996) showing limb-brightened dust-condensation zones.
Accordingly, the interpretation of the observational data by FWHM Gau\ss\
diameters may give misleading results.

The two components can be interpreted as if
\object{IRC\,+10\,420} has suffered from much higher mass-loss rates
in its recent past than today. For instance, the $T_{1}=1000\,$K model
gives
$\dot{M}_{1}= 1.4 \cdot 10^{-5}$\,$M_{\odot}/{\rm yr}\cdot (d/{\rm kpc})$
and
$\dot{M}_{2}= 8.5 \cdot 10^{-4}$\,$M_{\odot}/{\rm yr}\cdot (d/{\rm kpc})$.
The kinematic age
of the outer component gives a corresponding timescale of $\sim 100$\,yr
(for $d=5$\,kpc). If  $T_{2}=1200\,$K both shells are located closer to the
central star by approximately 30\%
leading to a correspondingly smaller timescale.
The failure of constant mass-loss wind models to fit the SED
agrees with the findings of Oudmaijer et al.\ (1996) and
Humphreys et al.\ (1997).
A previous high mass-loss episode 
is in line with the suspected post-RSG stage of
\object{IRC\,+10\,420}. 

Although, the present observations give only marginal evidence for
deviation from spherical symmetry
(if elliptical,
position angle of the long axis $\sim 130\degr \pm 20\degr$, 
axis ratio $\sim 1.0$ to 1.1),
the hot shell may also be interpreted
as a disk with a typical diameter of approximately 50\,mas.
The presence  of a rotating equatorial disk has been proposed 
by Jones et al.\ (1993), and  Oudmaijer et al.\ (1996)  interpreted their 
hot dust-shell  as a disk as well.
Provided the disk is not viewed pole-on, the corresponding
two-dimensional power spectra should be clearly elongated.
It should be noted, however, that 
disks with an extension of typically 50\,mas can only be detected 
in the power spectra if they provide at least, say, 10\%  of the total flux. 
Oudmaijer (1995) discussed several models for the circumstellar shell of 
\object{IRC\,+10\,420} and found neither a bipolar nor a disk-like  wind to
be consistent with optical and infrared high-resolution spectroscopy.
This seems to be supported by the present observations.
In order to be in line with optical
blue-shifted emission lines and red-shifted absorption lines
Oudmaijer suggested the scenario of infall of circumstellar 
material onto the stellar photosphere. However, 
according to Klochkova et al.\ (1997) the concept of accretion does not 
appear to be unproblematic either.

Thus, the question which scenario is best suited still
appears to be a matter of debate. 
Bispectrum speckle interferometry gives important information
on the spatial extension of the circumstellar shell.
It will be in particular the combination of different observations - 
photometry, spectroscopy and high-resolution imaging - and their simultaneous
modelling, which will  shed more light on the nature of
\object{IRC\,+10\,420} that is probably being witnessed in its transition
to the Wolf-Rayet phase.
%
%
\begin{acknowledgements}
The observations were made with  the SAO 6\,m telescope operated by the
Special Astrophysical Observatory, Russia.
We thank R.\ Oudmaijer for 
valuable discussions on  this outstanding object and
M.A.T.\ Groenewegen for providing us with the '1992' data set of the SED.
The radiative-transfer calculations are based on 
the code DUSTY developed by \v{Z}.\ Ivezi\'c, M.\ Nenkova and M. Elitzur.  
A.\ Gauger and A.\ Men'shchikov are thanked for 
various discussions on radiative transfer problems.
We are grateful to the referee for instructive remarks.
\end{acknowledgements}
\end{document}